
\input phyzzx.tex
\overfullrule0pt

\def\PRD#1#2#3{{\sl Phys.~Rev.} {\bf D#1}, #2 (#3)}
\def\PLB#1#2#3{{\sl Phys.~Lett.} {\bf B#1}, #2 (#3)}

\def\NPB#1#2#3{{\sl Nucl.~Phys.} {\bf B#1}, #2 (#3)}
\def\gev{{\rm  GeV }}
\def\fourth{\ifmath{{\textstyle{1\over 4}}}}
\def\suc{{\rm SU(3)_c}}
\def\sutwol{{\rm SU(2)_L}}
\def\hc{{\rm  h.c.}}
\def\mz{m_{\rm z}}
\def\calo{{\cal O}}
\def\msusy{M_{\rm SUSY}}
\def\tanb{\tan\beta}
\def\mhl{m_{h^0}}
\def\abs#1{\left|#1\right|}
\def\ls#1{\ifmath{_{\lower1.5pt\hbox{$\scriptstyle #1$}}}}
\def\ifmath#1{\relax\ifmmode #1\else $#1$\fi}

\def\refmark#1{~[#1]}
\def\hbar{{\overline H}}
\def\mhn{m_{N_1}}
\def\uy{{\rm U}(1)_{\rm Y}}
\def\uyp{{\rm U}(1)_{\rm Y^\prime}}
\def\n{${\rm \tilde n}$}

\date={}
\titlepage
\line{DESY 94-061\hfill ISSN 0418-9833}
\line{April 1994\hfill hep-ph/9404257}
\vskip1.cm

\title{\bf Can the Supersymmetric $\mu$
parameter be generated dynamically without
a light Singlet?}

\author{Ralf Hempfling}

\vskip .1in
\address{Deutsches Elektronen-Synchrotron, Notkestra\ss e 85,
D-22603 Hamburg, Germany}
\vfil
\abstract
It is generally assumed that the dynamical generation of the
Higgs mass parameter of the superpotential, $\mu$,
implies the existence of a light
singlet at or below the supersymmetry breaking scale, $\msusy$.
We present a counter-example in which the singlet field
can receive an arbitrarily heavy mass (\eg, of the order of the Planck scale,
$M_{\rm P}\approx 10^{19}~\gev$).
In this example, a non-zero value of $\mu$ is generated
through soft supersymmetry breaking parameters and is thus naturally of the
order of $\msusy$.
\endpage

\REF\susyrev{H.P. Nilles, {\sl
Phys.~Rep.} {\bf 110}, 1 (1984);
H.E. Haber and G.L.
Kane, {\sl Phys.~Rep.} {\bf 117}, 75 (1985);
R. Barbieri, {\sl Riv. Nuovo Cimento} {\bf 11}, 1 (1988).}%
\REF\lowmu{%
J.-F. Grivaz, in \sl Proceedings of the Workshop on $e^+e^-$ Collisions at
$500~\gev$: The Physics Potential, \rm Munich, Annecy, Hamburg, DESY report
DESY 92-123B (1992).}%
The cancellation of quadratic divergences in the unrenormalized Green
functions is one of the main motivations of supersymmetry (SUSY).
It stabilizes any mass scale under radiative corrections and thus allows the
existence of different mass scales such as the electroweak scale,
given by the $Z$ boson mass, $\mz$, and the
Planck scale, $M_{\rm P}$.
The minimal supersymmetric standard model (MSSM) is the most popular
model of this kind due to its minimal particle content\refmark\susyrev.
In this model, the SU(2)$_{\rm L}\otimes$U(1)$_{\rm Y}$
symmetry breaking is driven by soft SUSY breaking parameters.
Thus, the SUSY breaking scale, $\msusy$, has to be at or
slightly above $\mz$.
For this mechanism to work it is also necessary that
the SUSY Higgs mass parameter, $\abs{\mu} \lsim \msusy$.
This parameter also determines the chargino and neutralino mass spectrum.
 From here one can deduce a
experimental lower bound from LEP experiments of $\abs{\mu} \gsim \mz/4$
independent of $\tanb$\refmark\lowmu.
The fact that in the MSSM the $\mu$-parameter, which is a priori arbitrary,
has to lie within the narrow range
$$
\fourth\mz\lsim\abs{\mu}\lsim\msusy\,,\eqn\murange
$$
has been considered a problem of fine-tuning.
\REF\refmu{%
G.F. Guidice and A. Masiero, \PLB{206}{480}{1988};
J.E. Kim and H.P. Nilles, \PLB{263}{79}{1991};
J.A. Casas and C. Mu\n oz, \PLB{306}{288}{1993}.}
\REF\ssinglet{%
E. Witten, \PLB{105}{267}{1981};
L. Ib\'a\n ez and G.G. Ross, \PLB{110}{215}{1982};
P.V. Nanopoulos and K. Tamvakis, \PLB{113}{151}{1982}.}%
\REF\rosz{%
see, \eg, J. Ellis, J.F. Gunion, H.E. Haber, L. Roszkowski and F. Zwirner,
\PRD{39}{844}{1989}.}%
Possible attempts to try and solve this problem
are the inclusion of gravitational couplings\refmark\refmu\
or the introduction of additional fields\refmark\ssinglet.

The introduction of a singlet, $N_1$, under the
SU(3)$_{\rm c}\otimes$SU(2)$_{\rm L}\otimes$U(1)$_{\rm Y}$
standard model (SM) gauge group is the
most economical extension of the MSSM in which
eq.~\murange\ is natural\refmark\ssinglet.
Here, the $\mu$ parameter is generated dynamically:
$\mu = \lambda\vev{N_1} \neq 0$, where $\vev{N_1}$ is
the vacuum expectation value (VEV) of $N_1$.
In this model, one can achieve that the potential vanishes in
the direction of $N_1$ in the SUSY limit
by imposing a discrete symmetry.
If one includes soft SUSY breaking terms
then $N_1$ acquires a VEV and a mass of order of $\msusy$.
Thus, one inevitable consequence of
this mechanism is the presence of a singlet field under the SM gauge group
in the low energy theory. This leads to a severe loss of predictability
of the SUSY Higgs sector.
In particular, the MSSM prediction
$$
\mhl \leq \mz + \hbox{radiative corrections}\,,\eqn\mhllimit
$$
will be evaded\refmark\rosz.

\REF\hhg{see, \eg, J.F. Gunion, H.E. Haber, G.L. Kane, and S.
Dawson, \it The Higgs  Hunter's Guide, \rm (Addison-Wesley, Redwood City,
CA, 1990).}%
We will demonstrate in the following that it is also possible
to make $N_1$ heavy [say $\mhn = \calo(M_{\rm P})$]
while keeping $\vev{N_1} = \calo(\msusy)$ without fine-tuning.
In this limit we recover the predictive Higgs sector of the MSSM\refmark\hhg\
with its well defined upper limit of the lightest Higgs boson mass
[eq.~\mhllimit].

First we need to extend the symmetry group of our Lagrangian in order
to forbid the explicit Higgs mass term of the
superpotential, $W_H = \mu H\hbar$. We choose a continuous symmetry
which has to be gauged to avoid a massless Goldstone boson.
Our extended gauge group is
SU(3)$_{\rm c}\otimes$SU(2)$_{\rm L}\otimes$U(1)$_{\rm
Y}\otimes$U(1)$_{\rm Y^\prime}$.
Let us now consider a toy model with three singlets,
$N_i\sim (1,1,0,Y_i)$ where $Y_i = 2,-2,-1$ for $i = 1,2,3$.
Here, the first two numbers indicate the multiplicity
of $N_i$ under $\suc$ and $\sutwol$, and the third and fourth number denote
the charges under $\uy$ and $\uyp$.
The superpotential of this model is
$$
W_N = m N_1 N_2 - \lambda N_1 N_3^2\,.\eqn\blabla
$$
 From here we can derive the SUSY potential,
$V_{\rm SUSY} = V_F + V_{D^\prime}$, where
$$\eqalign{
&V_F = \abs{m N_2 - \lambda N_3^2}^2
+ \abs{m N_1}^2 + 4\abs{\lambda N_1 N_3}^2\,,\cr
&V_{D^\prime} = {g^{2\prime}\over 8}\left(\xi + 2 N_1^*N_1
- 2 N_2^*N_2 - N_3^*N_3\right)^2\,.}
\eqn\blabla
$$
\REF\fayetili{%
P. Fayet and J. Iliopoulos, \PLB{51}{461}{1974}.}%
\REF\oraif{%
\eg, L. O'Raifeartaigh, \NPB{96}{331}{1975}.}%
Here the inclusion of a Fayet-Iliopoulos term\refmark\fayetili, $\xi$,
is the easiest way of breaking the U(1)$_{\rm Y^\prime}$ gauge symmetry
but one can envisage other alternatives\refmark\oraif.
The VEVs are denoted by
$$\eqalign{
&n_1 = \vev{N_1} = 0\,,\cr
&n_2 = \vev{N_2} = \fourth\left(-{m\over\lambda}
+ \sqrt{{m^2\over\lambda^2} +4\xi}\right)\,,\cr
&n_3 = \vev{N_3} = \sqrt{m n_2\over\lambda}\,.\cr}
\eqn\blabla$$
The CP-even and CP-odd components of the scalar field $N_1$ are
mass-degenerate mass-eigenstates with $\mhn = (m^2+\lambda^2 n_3^2)^{1/2}$.
The gauge boson, $g^\prime$, acquires a mass
$m_{g^\prime} = g^\prime(n_2^2+n_3^2/4)^{1/2}$ via the Higgs
mechanism. The masses of the remaining CP-even (CP-odd) scalars are
$\mhn$, $m_{g^\prime}$
($\mhn$, $0$; the zero mass eigenvalue corresponds to the Goldstone
boson which is absorbed to give mass to the gauge
boson). The mass eigenvalues of the fermionic components are
$\pm\mhn$ and $\pm m_{g^\prime}$ as required if SUSY is unbroken.

\REF\rsymm{%
P. Fayet, \NPB{90}{104}{1975};
A. Salam and J. Strathdee, \NPB{87}{85}{1975}.}%
\REF\gira{L. Girardello and M.T. Grisaru,
{\sl Nucl. Phys.} {\bf B194}, 65 (1982).}%
Note that in addition to the gauge and the SUSY transformations the
Lagrangian is invariant under the global U(1) $R$-symmetry\refmark\rsymm\
which does not commute with SUSY.
This symmetry transforms
$\Phi \to \exp(i n_\Phi\alpha)\Phi$,
where $n_\Phi = 2,0,0,0$ for the bosons
and $n_{\widetilde\Phi}=1,-1,-1,1$ for the fermions
($\Phi = N_1,N_2,N_3,g^\prime$).
We now break SUSY explicitly in the standard fashion by including
soft SUSY breaking terms\refmark\gira
$$
V_{\rm soft} = Bm N_1 N_2 - A\lambda N_1 N_3^2+\hc\,, \eqn\softsusy
$$
where $A,B = \calo(\msusy)$ are the soft SUSY breaking parameters.
With these terms the $R$-symmetry is broken down to a discrete $Z_2$ symmetry
($\alpha = \pm \pi$).
If we minimize the full potential, $V = V_{\rm SUSY} +V_{\rm soft}$, we find
$$
\vev{N_1} \approx (A-B){m n_2\over \mhn^2}= \calo(\msusy) \neq 0\,.
\eqn\blabla
$$

We have seen that in our toy-model all the fields acquire a mass of the order
of the arbitrary U(1)$_{\rm Y^\prime}$ breaking scale parameterized by
$\sqrt{\xi}$.
However, the condition $\vev{N_1}= 0$ is protected by $R$-symmetry
to all orders in perturbation theory and is only
broken by adding soft SUSY breaking terms [eq.~\softsusy].
We now include in our model the full particle content
of the MSSM. The $Z_2$ symmetry is equivalent to the usual
$R$-parity that prevents baryon and lepton number violating interactions.
The full superpotential can then be written as
$$
W = W_N + W_H + W_Y\,,
\eqn\superpot$$
where $W_H = \kappa N_1 H \hbar$ and $W_Y$ are the standard Yukawa couplings.
The $Y^\prime$ assignments of the quark, leptons and Higgs particles
is constrained by the terms of $W$ in eq.~\superpot\ and by requiring the
absence of anomalies. These constraints can be satisfied
by introducing additional pairs of SUSY multiplets
$T\sim (n_c,n_w,Y,Y^\prime_1)$ and
$T^c\sim (\bar n_c,n_w,-Y,Y^\prime_2)$.
These representations have been included in pairs such that
below the U(1)$_{\rm Y^\prime}$ breaking scale, $\sqrt{\xi}$,
the dynamic mass terms
$W\sim  m_T T T^c$ can arise ($m_T = \vev{N_i}^n/\xi^{(n-1)/2}$;
$n = 1,2,..$ and $i = 1,2,3$).
If we assume that all three generations have the same
$\uyp$ charges then the absence of anomalies requires the existence
of at least one pair of color non-singlets, $T$ and
$T^c$ for which $Y^\prime_1+Y^\prime_2>0$ and thus $m_T \propto \vev{N_1} =
\calo(\msusy)$. However, the Higgs particles present
at $\msusy$ and thus also the Higgs couplings are equivalent to the MSSM.
\ack{Useful conversations with W. Buchm\"uller are gratefully acknowledged}
\refout
\end

We have presented a model in which the condition
$\mu = \calo(\msusy)$ arises naturally.
This model reduces in the low energy limit to the MSSM without any
SU(2)$_{\rm L}\otimes$U(1)$_{\rm Y}$ singlets.
The necessary ingredient for this mechanism is the existence of
a global symmetry (in our case a U(1) symmetry) of the Lagrangian
which does not commute with SUSY and is broken by soft SUSY breaking terms.